\documentclass[prl,aps,superscriptaddress,10pt,nobibnotes,notitlepage,twocolumn,nofootinbib,preprintnumbers]{revtex4-2}

\usepackage[hyperindex,breaklinks,colorlinks=true,citecolor=blue,hyperfootnotes=false]{hyperref}
\usepackage[english]{babel}
\usepackage{amssymb,amsmath}
\usepackage{mathrsfs}
\usepackage[dvipsnames,x11names]{xcolor}
\usepackage{graphicx}
\usepackage{mathtools}
\usepackage{empheq}
\usepackage{marginnote}
\usepackage[cal=cm,scr=euler]{mathalpha}
\usepackage{cleveref}
\usepackage{cancel}

\begin{document}

\title{Quantum and Thermal Fluctuations of Cherenkov Radiation from HQET}

\def\ARGaffil{Physics Division, Argonne National Laboratory, Lemont, Illinois, 60439, U.S.A}
\def\KITPaffil{Kavli Institute for Theoretical Physics, University of California, Santa Barbara, California 93106, USA}

\author{Joshua Lin}
\email{joshua.lin@anl.gov}
\affiliation{\ARGaffil}
\author{Bruno Scheihing-Hitschfeld}
\email{bscheihi@kitp.ucsb.edu}
\affiliation{\KITPaffil}

\date{\today}

\begin{abstract}
Charged particles travelling faster than the speed of light in the medium in which they propagate 
emit Cherenkov radiation. The formula for the spectrum of this radiation as a function of frequency, known as the Frank-Tamm formula, first derived almost 90 years ago~\cite{Frank:1937fk}, follows purely from classical electromagnetism. 
In this work, we demonstrate how this result also follows from a short quantum field theory calculation, which in addition to it contains all of the cumulants of thermal and quantum fluctuations around the classical radiation spectrum at leading order in the inverse of the particle's mass. All of these results follow from the particle's momentum change probability, which we calculate for weakly coupled gauge theories using the tools of Heavy Quark Effective theory. 
\end{abstract}

\maketitle

It is well-known that the classical behaviour of 
charged particles travelling faster than the speed of light in the medium in which they propagate is that they emit Cherenkov radiation. In its simplest form, the celebrated Frank-Tamm formula~\cite{Frank:1937fk}: 
\begin{equation}
    \frac{\partial^2 E}{\partial x \partial \omega } = \frac{q^2 \omega}{4\pi c_{\rm med} }  \left( 1 - \frac{c_{\rm med}^2}{v^2} \right) \theta(v - c_{\rm med}) \, , \label{eq:Frank-Tamm}
\end{equation}
describes the rate at which such a particle radiates energy into the medium per unit of path length $x$ and frequency $\omega$. Here, $q$ is the electric charge of the particle, and $c_{\rm med}$ is the speed of light in the medium.\footnote{We use units where the reduced Planck's constant is equal to unity $\hbar = 1$ and the electric charge is dimensionless.} For simplicity, we have omitted the frequency dependence of material properties in this equation. As the Heaviside step function indicates, the radiated energy vanishes if $v < c_{\rm med}$.

By comparison to other radiative phenomena, this is a particularly simple phenomenon, because it is generic:  its occurrence only depends on the ratio between the speed at which the particle propagates and the speed of propagation of signals in the medium. 

On the other hand, the origin of radiative phenomena and energy loss of charged particles is intrinsically tied to quantum mechanics~\cite{bohr1948penetration,Jackson:1998nia}. However, the available derivations of the Cherenkov radiation spectrum focus chiefly on the average energy loss experienced by the charged particle, with both classical and quantum methods yielding equivalent answers~\cite{Ginzburg_1996}. 
A straightforward, self-contained quantum field theory derivation of the Cherenkov radiation spectrum with all of its quantum fluctuations can thus provide new insight into the physics of radiation and energy loss. In addition to that, it would provide the possibility to interface with modern developments in high-energy physics in the same language and provide a baseline from which systematic improvements can be carried out using these tools. Our purpose here will be less about establishing new applications, but rather to use Cherenkov radiation as a conceptually clean probe of our understanding of energy loss in high-energy physics.
That being said, applications of Cherenkov radiation in high energy physics are plentiful, from the extremely practical in the functioning of detectors in collider experiments~\cite{methods-detection,GEANT4:2002zbu}, through a long history of research into the possibility of Cherenkov emission in nuclear matter~\cite{Wada:1949zz,Ivanenko1949,Blokhintsev1950,Yekutieli1959,Czyz1959,Czyz1960,Smrz1962,Ion1970,Ion1971,Ion1971thesis,Zaretskii1977,Dremin1979,Dremin1981,Dremin:2005an,Price:2026mof}, to the conjectural 
as signatures of new physics in UV extensions of the Standard Model that break Lorentz invariance, e.g.~\cite{Kaufhold:2005vj,Altschul:2006zz,Cheng:2006us,Carroll:2009mr,DeLaurentis:2012fh} 
(which have in turn been strongly constrained by observation~\cite{Kimura:2011qn,Yagi:2013ava,Yagi:2013qpa,Yunes:2016jcc}).

In this Letter, we calculate the complete statistics of the thermal and quantum fluctuations of the Cherenkov radiation spectrum at leading order in $1/M$, with $M$ the mass of the charged particle, starting from first-principles quantum field theory. We do so by leveraging techniques provided by Heavy Quark Effective Theory (HQET), of which we give a brief overview in what follows. Once all of its relevant ingredients are laid out, the derivation of the Frank-Tamm formula as the mean of a probability distribution follows straightforwardly. Interestingly, the full statistics of the radiation spectrum is far from Gaussian, and it shows significant asymmetries between radiating more and less energy than the mean, which we shall describe in detail.  

Because it is based on HQET, at the formal level the starting point of our calculation is the same as that of recent theoretical studies of heavy quark energy loss in quark-gluon plasma (QGP) and related Yang-Mills theories~\cite{Rajagopal:2025ukd,Rajagopal:2025rxr}.
Experimentally, such effects have been measured in the suppression of $D$- and $B$-meson production as compared to $pp$ collisions \cite{CMS:2017qjw,ALICE:2018lyv,ALICE:2021mgk,ALICE:2021rxa,PHENIX:2022wim,CMS:2024vip}. 
However, understanding the exact mechanism behind this energy loss is difficult not only because of the need to connect quark/gluon degrees of freedom in terms of which calculations are carried out and the hadronic degrees of freedom measured experimentally, but also simply due to the theoretical difficulty in calculating the expected energy loss. 
This has motivated a lot of research over the years (for reviews see~\cite{Baier:2000mf,Casalderrey-Solana:2007knd,Rapp:2018qla,Wang:2025lct}), 
including by considering the possibility that energy loss proceeds via Cherenkov radiation~\cite{Koch:2005sx,Casalderrey-Solana:2009zbk,Casalderrey-Solana:2010cfu}.
While we do not seek to make direct contact with heavy-ion phenomenology in this work, our results are demonstrative of a simple, yet powerful feature of the HQET formulation we utilize: at no point there is an assumption of what the energy loss mechanism at play is. Whether radiative or collisional, the Wilson loop we shall discuss contains, from its formulation, all mechanisms relevant for such a heavy particle.
{Our results demonstrate via an explicit example how the interplay between zero temperature quantum effects and finite temperature effects can take place at the level of the momentum change probability, which is essential in order to further our understanding of such phenomena.}

\textbf{Energy loss and momentum change of test particles in gauge theories.}
In the framework of HQET~\cite{Georgi:1990um}, the physics of small momentum transfer imparted on a heavy fermion of mass $M$ moving in spacetime with 4-velocity $v$ is described by the Lagrangian density:
\begin{equation}
    \mathcal{L}_{\rm HQET} = \bar{Q} (i v \cdot D) Q + \mathcal{L}_{\rm light} + \mathcal{O}(1/M) \, , \label{eq:HQET}
\end{equation}
where $Q = Q(x)$ is the field describing the small momentum fluctuations $k^\mu$ around the ``hard'' heavy particle momentum $M v^\mu$. That is to say, the total heavy particle momentum is $p^\mu = M v^\mu + k^\mu$, where $k^\mu$ is the momentum variable conjugate to the position argument of the field $Q(x)$. At leading order in $1/M$, the gauge field in the covariant derivative $D_\mu = \partial_\mu - ig A_\mu$ gives rise to all of the nontrivial dynamics of the heavy particle, as all of the heavy particle's interactions with the light degrees of freedom of the theory -- encoded in $\mathcal{L}_{\rm light}$ -- are accounted by this term.

With minimal modifications, this Lagrangian is broadly applicable to the description of heavy charged particles coupled to a gauge field, regardless of their gauge group representation and other quantum numbers. 
The dynamics stemming from the Lagrangian~\eqref{eq:HQET} does not change the spacetime direction in which the particles propagate. Rather, all it does is to parallel transport them along straight lines characterized by their velocity $v^\mu$ --- the effects of
which are only dependent on properties of the gauge theory describing the light degrees of freedom to which the particle is coupled.
This can be summarized in a relatively simple expression for the quantum mechanical transition amplitude between two heavy particle states whose 3-momentum differs by an amount $\Delta {\bf k}$:
\begin{equation}
    \langle {\bf p} - \Delta {\bf k} |_a U_{\rm HQET}(t_f-t_i) | {\bf p} \rangle_b = \int \mathrm{d}^3{\bf x}_f \ e^{i \Delta {\bf k} \cdot {\bf x}_f } W^{ab}_{[x_f,x_i]} \, , \label{eq:amplitude}
\end{equation}
where $U_{\rm HQET}$ is the time evolution operator that the Lagrangian in Eq.~\eqref{eq:HQET} defines, $|{\bf p}\rangle_a$ is a heavy particle state with momentum ${\bf p}$ and color $a$, and $W$ is the Wilson line 
\begin{equation}
    W^{ab}_{[x_f,x_i]} = \mathrm{Pexp} \left( i g \int_{t_i}^{t_f} \mathrm{d}t \,  A^A_{\mu} T^A_{ab} \dot{x}^\mu \right) \, \label{eq:wilson-line}
\end{equation}
along the path $x^\mu(t) = x_i^\mu + (t-t_i, {\bf v} (t-t_i))$ up to $x_f = x^\mu(t_f)$. 

At the purely quantum level, Eq.~\eqref{eq:amplitude} contains all the information we could ask about the dynamics of the heavy particle, for any gauge field configuration $A_\mu$ (which insofar only serves as a background on which the particle propagates). 
Our main focus in this work is the gauge field state defined by a thermal ensemble with density matrix $\propto e^{-\beta H}$, 
{with $H$ the Hamiltonian of the environment degrees of freedom, i.e., the Legendre transform of $\int \mathrm{d}^3x \mathcal{L}_{\rm light}$}.
As such, we introduce an initial state described by a density matrix given by
\begin{equation}\label{eq:factorised-dressing}
    \rho =
    |{\bf p}\rangle_a e^{-\beta H} \rho_{ab}^{Q} \langle {\bf p}|_b \, ,
\end{equation}
representing a dressed heavy quark state inserted on top of the light thermal medium. The dressing factor $\rho_{ab}^Q$ encodes the initial state preparation of the color degrees of freedom of the heavy quark, which must be suitably entangled with the gauge field degrees of freedom (e.g., via a Wilson line) to define a gauge-invariant density matrix $\rho$\footnote{Technically, the dressing cannot be written in a way that factorises in \Cref{eq:factorised-dressing}. For example, a simple Hermitian dressing is given by $\int \mathrm{d}^3\mathbf{x} \mathrm{d}^3 \mathbf{y} e^{i \mathbf{p}\cdot(\mathbf{x}-\mathbf{y})}  \ Q^\dagger(x) \psi (x) \ e^{-\beta H_\mathrm{light}} \ \psi^\dagger(y) Q(y) $ for situations where there are light fermions $\psi$ in the environment.  
This does not affect the later calculations however, which are all carried out in the large $t$ limit.}. {In Quantum Electrodynamics (QED), the color indices $a,b$ are absent.} 

It is then natural to calculate the probability that said heavy particle evolves into a state with momentum ${\bf p} - \Delta{\bf k}$ after a time $t$ has passed, which one can write as 
\begin{align}\label{eq:alignsome}
    &P(\Delta {\bf k};{\bf v}({\bf p}))=\nonumber \\
    &\frac{{\rm Tr} \! \left[ |{\bf p} - \Delta {\bf k}\rangle_{c} \tilde{\rho}^Q_{cd} \langle {\bf p}-\Delta{\bf k}|_d U(t) |{\bf p}\rangle_a e^{-\beta H} \rho_{ab}^{Q} \langle {\bf p}|_b U^\dagger(t) \right]}{{\rm Tr} \! \left[ U(t) |{\bf p}\rangle_a e^{-\beta H} \rho_{ab}^{Q} \langle {\bf p}|_b U^\dagger(t) \right]} \, ,
\end{align}
where $\mathbf{v}(\mathbf{p}) = {\bf p}/\sqrt{M^2 + {\bf p}^2}$ denotes the 3-velocity of the heavy particle, which reparametrizes the $v$ parameter of HQET 
and $\tilde{\rho}^Q$ defines the dressing of the final state projection to make the trace gauge-invariant. The trace goes over all physical states of the theory.
We will keep track of the momentum change of the heavy particle 
but its velocity will not change because ${\bf v}({\bf p} - \Delta {\bf k}) = {\bf v}({\bf p}) + \mathcal{O}(\Delta{\bf k}/M)$, and we assume ${\Delta \bf k}/M \ll 1$ throughout.
Note that the denominator of \Cref{eq:alignsome} is actually time-independent, because the time evolution operators compensate each other.

It is then straightforward to use the HQET amplitude~\eqref{eq:amplitude} to arrive at a more explicit expression for the momentum transfer probability to a heavy particle propagating through a thermal environment. After a short calculation one obtains 
\begin{align}
    P(\Delta {\bf k};{\bf v}) &= \frac{1}{(2\pi)^3} \int \mathrm{d}^3{\bf L} \, e^{-i \Delta {\bf k} \cdot {\bf L}  } \langle W_{\bf v} \rangle ({\bf L}) \, , \label{eq:P-from-W}
\end{align}
where $ \langle W_{\bf v} \rangle ({\bf L})$ is an expectation value of a Wilson loop
\begin{align}
    &\langle W_{\bf v} \rangle ({\bf L}) = \nonumber \\ 
    &\ \ \ \frac{1}{Z} {\rm Tr}_{\mathcal{H}} \! \left[ W^{bc}_{[(0, {\bf L}),(t, {\bf v} t + {\bf L} )]} \tilde{\rho}_{cd}^Q W^{da}_{[(t, {\bf v} t ),(0,{\bf 0})]} e^{-\beta H}  \rho_{ab}^Q \right] \, ,
    \label{eq:Wloop-Wlines}
\end{align}
which does not rely on any perturbative expansion or approximation other than the large mass required to use the HQET Lagrangian~\eqref{eq:HQET}. In practice, this sets a parametric limitation on the momentum transfer ($|\Delta {\mathbf{k}}| \ll M$ in the rest frame of the heavy particle) 
for this calculation to reproduce the result of the complete theory.

In the large time 
limit, time-translation invariance of the theory implies that the time dependence of the Wilson loop is exponential
\begin{equation}
    \langle W_{\bf v} \rangle ({\bf L}) = \exp \left( - t S({\bf L}; {\bf v}) + \ldots \right) .
    \label{eq:W-S}
\end{equation}
Note that $S$ is insensitive to the details of the initial and final color components of the heavy particle state. Physically, this is a consequence of the color degrees of freedom of the particle having a long time to become (effectively) randomized via interactions with the environment. Mathematically, it is a consequence of the fact that the $t$-extensive part of the logarithm of the Wilson loop is solely determined by the correlations between the gauge fields in the two long antiparallel Wilson lines in Eq.~\eqref{eq:Wloop-Wlines}. The initial and final state projections contribute only to the $\mathcal{O}(1)$ prefactor (i.e., non-extensive in $t$) in Eq.~\eqref{eq:W-S}.

For purposes of studying the late-time limit of $P(\Delta \mathbf{k},\mathbf{v})$, 
a crucial result is that regardless of the details of the theory, this function satisfies
\begin{equation}
    S({\bf L};{\bf v}) = S(-{\bf L} + i {\bf v}/T;{\bf v}) \, , \label{eq:S-analytic-property}
\end{equation}
as a consequence of a generalized KMS relation between line operators~\cite{Rajagopal:2025rxr}.
This relation \textit{enforces} that the momentum kicks provided by the medium are asymmetric:
\begin{equation}
    P(\Delta \mathbf{k},\mathbf{v}) = P(-\Delta \mathbf{k},\mathbf{v}) \mathrm{exp} \left(\frac{\mathbf{v} \cdot \Delta \mathbf{k}}{T} \right) \label{eq:asymm-P}
\end{equation}
and in the long time limit around $\Delta \mathbf{k} \sim 0$:
\begin{equation}
\lim_{t \to \infty} P(\Delta \mathbf{k},\mathbf{v}) \propto \mathrm{exp} \left(\frac{\mathbf{v} \cdot \Delta \mathbf{k}}{2T} \right) \label{eq:late-time}
\end{equation}
which, when solving the kinetic equation for the evolution of the heavy particle phase space distribution $f(\mathbf{k},t)$ that follows from $P$ and Eq.~\eqref{eq:W-S}  
\begin{equation}
    {\partial_t f({\bf k},t) = - S(-i\partial_{\bf k} ; {\bf v}) f({\bf k},t) \, ,} \label{eq:kinetic}
\end{equation}
provides a steady state that is given by the appropriate thermal state according to the linearised dispersion relation:
\begin{equation}\label{eq:thermal}
f_\mathrm{thermal}(\mathbf{k}) \propto \mathrm{exp}\left(-\frac{\mathbf{v}\cdot\mathbf{k}}{T}\right) \, .
\end{equation}
Notice that the thermal state in \Cref{eq:thermal} is un-normalisable, as we have restricted in this case to a single HQET $v$-sector and the equations are only valid for $|\mathbf{k}| \ll M$. However, if one allows for changes in the heavy particle momentum that cumulatively change its velocity, as discussed in Ref.~\cite{Rajagopal:2025rxr}, one can show that the Boltzmann distribution $\propto \exp(-E({\bf p}/T))$ with ${\bf v} = \partial E/\partial {\bf p}$ satisfies the detailed balance condition as a stationary distribution by virtue of  Eq.~\eqref{eq:asymm-P}. 

Whether this steady state can be reached dynamically is a different question. In fact, for the setup we consider here, there will always be a net flow of particles from higher momentum towards lower momentum. If we allowed for their velocity to change, after some time they would start to accumulate at or slightly below $v = c_{\rm med}$, and as they reach this point they would stop evolving from then on because $P(\Delta {\bf k}; |{\bf v}| < c_{\rm med}) = \delta(\Delta {\bf k})$. 

\textbf{Cherenkov radiation.}
In a Gaussian field theory (or in the weakly coupled limit of interacting theories), this Wilson loop is determined by the Wightman function $D^>_{\mu \nu}$ of the gauge field
\begin{align} \label{eq:W-loop-wightman-general}
    &\langle W_{\bf v} \rangle ({\bf L}) = \\
    & N \exp \!\left( \!g^2 C_2(R) n^\mu n^\nu \!\!\! \int_{0}^t \! \! \! \mathrm{d}t' \mathrm{d}t'' D_{\mu \nu}^{>}(t'-t'',{\bf v} (t'-t'') + {\bf L} ) \!\right) \nonumber
\end{align}
where $C_2(R)$ is the quadratic Casimir of the gauge representation of the heavy particle $R$, $n^\mu = (1, {\bf v})$ is the direction that characterizes the heavy particle's trajectory, $N$ is an overall ${\bf L}$-independent constant (but possibly dependent on $t$ and ${\bf v}$) enforcing $\langle W_{\bf v} \rangle ({\bf 0}) = 1$, and the Wightman function is
\begin{align}
    D_{\mu \nu}^{>}(t,{\bf x}) = \langle A_\mu(t,{\bf x}) A_\nu(0,{\bf 0}) \rangle_T \, . \label{eq:D-bigger}
\end{align}
{In fact, Eqs.~\eqref{eq:W-loop-wightman-general} and~\eqref{eq:D-bigger} provide the \textit{exact} answer for QED, provided all the effects of interactions with the light fermionic degrees of freedom are included in the Wightman function.} 

{In the long-time limit,
\begin{align}
    \int_{0}^t \! \mathrm{d}t' \mathrm{d}t'' D_{\mu \nu}^{>}(t'-t'',{\bf v} (t'-t'') + {\bf L}) \nonumber \\ \overset{t \to \infty}{\longrightarrow}  t \int_{-\infty}^\infty \!\!\!\! \mathrm{d}t' D_{\mu \nu}^{>}(t',{\bf v} t' + {\bf L} ) \, , \label{eq:limit}
\end{align}
and the formal expression~\eqref{eq:W-S} becomes explicit, with}
\begin{align}
    S({\bf L}; {\bf v}) &= - g^2 C_2(R) n^\mu n^\nu \\ & \quad \times \int_{-\infty}^\infty \mathrm{d}t' \left[ D_{\mu \nu}^{>}(t',{\bf v} t' + {\bf L} ) - D_{\mu \nu}^{>}(t',{\bf v} t' )\right] \, . \nonumber
\end{align}
{where the neglected ``$\ldots$'' terms in Eq~\eqref{eq:W-S} are controlled by how quickly the limit~\eqref{eq:limit} converges (in turn controlled by the energy scales that enter the Wightman function).}
This expression automatically satisfies Eq.~\eqref{eq:S-analytic-property} because the Wightman correlator satisfies $D^>(t,{\bf x}) = D^>(-t-i\beta,-{\bf x})$ in thermal equilibrium. The last term inside the square brackets ensures that $S({\bf 0}; {\bf v}) = 0$ and therefore $\langle W_{\bf v} \rangle ({\bf 0}) = 1$, thus defining a normalized momentum change probability distribution~\eqref{eq:P-from-W}.

{At zeroth order in a weak coupling expansion,} the gauge field Wightman correlator in Feynman gauge is given by: 
\begin{align}
    D^>_{\mu \nu}(\omega,{\bf k}) 
    &= -c_{\rm m}^3 g_{\mu \nu}\frac{ {\rm sgn}(\omega)  }{1- e^{-\omega/T}} (2\pi) \delta(\omega^2 - c_{\rm m}^2 {\bf k}^2 ) \, , 
\end{align}
where $g_{\mu \nu} = {\rm diag}(1,-c_{\rm m}^{-2},-c_{\rm m}^{-2},-c_{\rm m}^{-2})$, together with the dispersion relation $\omega = c_{\rm m}|{\bf k}|$ characterize the gauge field two-point function in a medium with subluminal propagation $c_{\rm m} < 1$. This gives $S(\mathbf{L};\mathbf{v})$ as {a numerically calculable integral (pending the introduction of cutoffs encoding material properties)}:
\begin{align}\label{eq:Sfinal}
&S(\mathbf{L},\mathbf{v}) =  \frac{g^2 {C_2} (c_{\rm m}^2-v^2) }{8 \pi^2 v c_{\rm m} }    \\
&\qquad \int_{-\infty}^{+\infty} \mathrm{d}\omega \int_0^{2\pi} \mathrm{d}\phi \frac{(e^{i L_\perp k_\perp \cos \phi } e^{i \omega L_\|/v}-1)\ \mathrm{sgn}(\omega)}{1 - e^{-\omega/T}} \nonumber \, ,
\end{align}
where $v = |{\bf v}|$, $k_\perp = \tfrac{\sqrt{v^2 - c_{\rm m}^2}}{vc_{\rm m}} |\omega| $, and $\mathbf{L} = \frac{L_\|\mathbf{v}}{v} + \mathbf{L}_\perp$ has been broken up into longitudinal and transverse components, and $\phi = \mathrm{arccos}\left( \frac{\mathbf{k_\perp \cdot L_\perp}\mathbf{}}{|\mathbf{k}_\perp||\mathbf{L}_\perp|}\right)$ is the relative angle between the perpendicular component of the gauge boson momentum, and the perpendicular component of $\mathbf{L}$. {We have assumed, as we do throughout the rest of this work, that $v > c_{\rm m}$.}

By using Eq.~\eqref{eq:Sfinal} in the kinetic equation~\eqref{eq:kinetic}, we get
\begin{align}
    \partial_t f({\bf q},t) 
    &= - \frac{g^2 C_2 (c_{\rm m}^2 - v^2) }{8\pi^2 v c_{\rm m} } \int_0^{\infty}\mathrm{d} \omega \int_0^{2\pi} \mathrm{d}\phi  \\ & \quad \left[ \frac{f({\bf q} + {\bf k},t) - f({\bf q},t)}{1 - e^{-\omega/T} } 
    + \frac{f({\bf q} - {\bf k},t) - f({\bf q},t)}{e^{\omega/T} - 1 } \right] \nonumber
\end{align}
and it becomes clear that the modes in Eq.~\eqref{eq:Sfinal} 
with $\omega > 0$ correspond to emission of gauge bosons (first term in the square bracket) whereas $\omega < 0$ corresponds to absorption of gauge bosons from the thermal medium (second term in square bracket).
Here ${\bf k} = (k_\|, {\bf k}_\perp)$ and $k_\| = \omega/v$. 
Hence, absorption can only take place if it is stimulated by the occupation number of the medium.

\textit{Rederivation of the Frank-Tamm formula.}
To characterise the dynamics generated by \Cref{eq:Sfinal}, note that for a given probability distribution $f(\mathbf{k},t)$ over $\mathbf{k}$ phase-space, the dynamics can be expanded in the Kramers-Moyal expansion form~\cite{Kramers1940,Moyal1949} as:
\begin{equation}
\frac{\mathrm{d}}{\mathrm{d}t} f(\mathbf{k},t) = \sum_{n=1}^\infty \frac{1}{n!} 
M_{i_1,\cdots,i_n}
\nabla_{i_1} \cdots \nabla_{i_n} f(\mathbf{k},t),
\end{equation}
where the derivatives $\nabla$ on the right hand side are taken with respect to $\mathbf{k}$, and {$M_{i_1,\cdots,i_n}$ are} the connected moments {of $P(\Delta {\bf k};{\bf v})$ per unit of time, given by:} 
\begin{equation}
M_{i_1,\cdots,i_n}
= -(-i)^n\  \nabla_{i_1} \cdots  \nabla_{i_n}S(\mathbf{L},\mathbf{v}) \big|_{\mathbf{L} = 0},
\end{equation}
where the derivatives are taken with respect to $\mathbf{L}$. 

Note that the first connected moment equivalently provides the information of the average momentum transfer from the heavy particle to the medium:
\begin{equation}
\frac{\mathrm{d}}{\mathrm{d} t} \int \mathrm{d}^3 \mathbf{k} \ f(\mathbf{k},t) \mathbf{k}_i  = - M_i,
\end{equation}
where the first moment is given by:
\begin{equation}
M_i = \mathbf{v}_i \left( \frac{g^2 C_2 (v^2-c_{\rm m}^2)}{4 \pi v^3 c_{\rm m} } \int_0^\infty \mathrm{d} \omega \ \omega \cdot \kappa(\omega) \right). 
\end{equation}
$\kappa(\omega)$ has been introduced as a regulator, for instance $\kappa(\omega) = \theta(\omega_\mathrm{UV} - |\omega|) \theta(|\omega| - \omega_\mathrm{IR})$ corresponds to a cutoff regulator where only modes with energy greater than $\omega_\mathrm{IR}$ but less than $\omega_{\mathrm{UV}}$ contribute. {In real materials, this encodes the physical statement that $c_{\rm m}$ is frequency-dependent and approaches the speed of light in vacuum at {small and }large frequencies.}
The KMS relation~\eqref{eq:S-analytic-property} holds `mode-by-mode,' {so it is unaffected by the regulator.} 

Note that the mean momentum transfer is purely in the longitudinal direction. Nonetheless, the angular distribution (uniform within a cone) of radiation into the medium is apparent from Eq.~\eqref{eq:Sfinal}.
Thus, the differential energy emitted per unit length travelled, is given by: 
\begin{equation}
\frac{\mathrm{d} E}{\mathrm{d} l} = \frac{\mathbf{v}_i}{v}  M_i  = \frac{g^2 C_2}{4\pi c_{\rm m}} \int \mathrm{d}\omega \ \omega \cdot \kappa(\omega) \left( 1 - \frac{c_{\rm m}^2}{v^2} \right) \, ,
\end{equation}
which is exactly the integral version of the Frank-Tamm formula~\eqref{eq:Frank-Tamm} for a general gauge group. 

\begin{figure*}[t!]
\centering
    \includegraphics[width=1.0\linewidth]{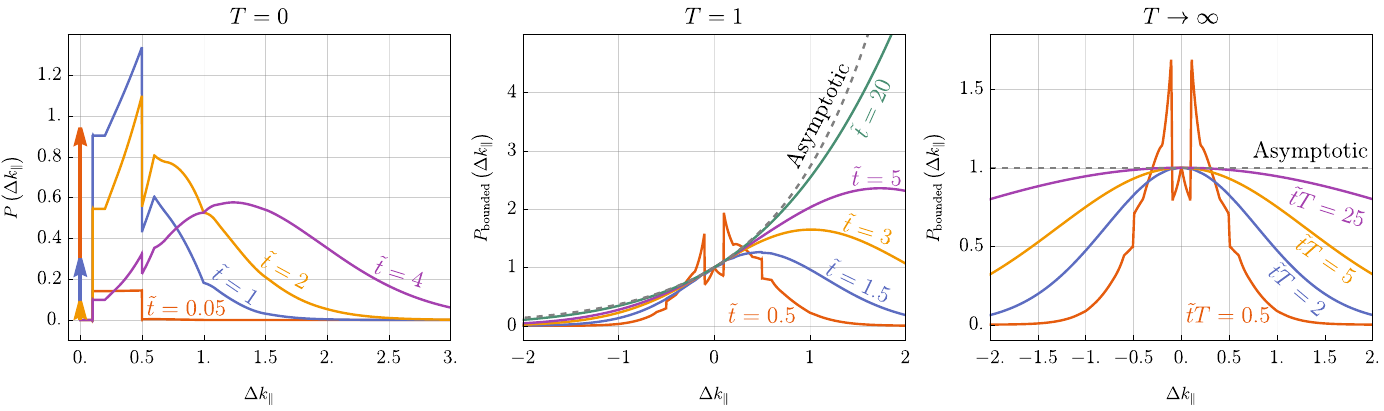}
    \vspace*{-0.75cm}
    \caption{Plots of $P(\Delta k_\|,v = 2c_m)$ for the hard cutoffs $\omega_\mathrm{IR} = 0.2, \omega_\mathrm{UV} = 1$. 
    The distributions at different normalised times $\tilde{t} = t \frac{g^2 C_2}{4 \pi^2}$ are plotted in different colours. 
    All the distributions have a component proportional to $\delta(k_\|)$ (\Cref{eq:delta}), these have been explicitly drawn for the $T = 0$ distributions as arrows --- for the $T = 1$ and $T \to \infty$ distributions however they are not shown as their norms are very small. 
    The asymptotic ($t \to \infty$) distributions for $T = 1$ and $T \to \infty$ are shown in dashed gray line, and the bounded part ($P_\mathrm{bounded}(\Delta k_\|)$) of the distribution have been rescaled such that $P_\mathrm{bounded}(0) = 1$. 
    }
    \label{fig:fig}
\end{figure*}

\textit{Quantum and thermal fluctuations.}
At zero temperature, higher moments of the distribution, {from the variance onward,} encode quantum fluctuations of the radiation pattern --- {the reason why there is a probability distribution, as opposed to a fully deterministic value of the energy loss, is precisely due to quantum effects}. If this process occurs at nonzero temperature, the magnitude of these fluctuations is also modified. One can straightforwardly show that the contribution to each connected moment (``cumulants'') of the longitudinal momentum fluctuations per unit time $M_\|^{(n)} \equiv M_{i_1=\|, \cdots , i_n=\|}$, and per unit of frequency, is given by 
\begin{align} 
    \frac{\mathrm{d} M_\|^{(n)} }{\mathrm{d}\omega} 
    = \frac{g^2 C_2 c_{\rm m} v}{4\pi}  \left( \frac{1}{c_{\rm m}^2} - \frac{1}{v^2} \right) \frac{\omega^n}{v^n} \frac{e^{\omega/T} + (-1)^n }{e^{\omega/T} - 1 } \, . 
\end{align}
{where the frequency element $\mathrm{d}\omega$ includes both the positive and negative frequency contributions of the same magnitude (i.e., it is a sum of absorption and radiation processes).}
{It is interesting to note that only the even cumulants differ from their vacuum counterparts. In particular, the mean of the momentum change distribution is unchanged by finite temperature effects. However, the variance, and all the other even cumulants, get enhanced as $T \to \infty$.

We remark on two salient properties of these non-Gaussian features:
\begin{enumerate}
    \item Because they follow from Eq.~\eqref{eq:Sfinal}, and the KMS condition~\eqref{eq:S-analytic-property} actually holds mode by mode in this expression, these cumulants per unit frequency 
    satisfy a sum rule for any $v$:
    \begin{equation}
        \sum_{n=1}^\infty \frac{(-1)^n}{n!} \frac{v^n}{T^n} \frac{\mathrm{d} M_\|^{(n)}}{\mathrm{d}\omega} = 0 \, ,
    \end{equation} 
    which links all moments of the (longitudinal momentum transfer) distribution.
    \item These cumulants are a consequence of quantum fluctuations of the light degrees of freedom in the medium. They are different from the quantum fluctuations coming from the particle's own motion, which appear as $1/M$ corrections in this HQET setup, and are suppressed in this large mass limit. 
\end{enumerate}
}

Plots of the longitudinal momentum transfer given by:
\begin{equation}
    P(\Delta k_\|,v) = \int \mathrm{d}^2 \Delta\mathbf{k}_\perp P(\Delta \mathbf{k},\mathbf{v})  
\end{equation}
are shown in \Cref{fig:fig} for the hard cutoff regulator $\kappa(\omega) = \theta(\omega_\mathrm{UV} - |\omega|)\theta(|\omega| - \omega_\mathrm{IR})$. 
Because all factors of time appear with corresponding factors of the gauge coupling, it has been relabelled $\tilde{t} = t \frac{g^2 C_2}{4 \pi}$. 
Explicit expressions for the longitudinal momentum transfer function in the $T \to 0$ and $T \to \infty$ limits are given by:
\begin{align}
&P(\Delta k_\|,v)_{T = 0} = e^{-\tilde{t} \frac{v^2 - c_{\rm m}^2}{v c_{\rm m}} (\omega_\mathrm{UV} - \omega_\mathrm{IR})} \bigg[ \delta(\Delta k_\|) + \int \frac{\mathrm{d} L_\|}{2 \pi}  \nonumber \\
&e^{-i k_\| L_\|} \!\!\left( e^{-\tilde{t} \frac{v^2 - c_{\rm m}^2}{v c_{\rm m}} \frac{1}{iL_\|/v} (e^{i L_\| \omega_\mathrm{IR}/v} - e^{i L_\| \omega_\mathrm{UV}/v})} - 1 \right)\bigg] 
\\
&P(\Delta k_\|, v)_{T \to \infty} = \int \frac{\mathrm{d}L_\|}{2 \pi} e^{-i \Delta k_\| L_\|}    \\
&e^{ -2\tilde{t}T\frac{v^2 - c_{\rm m}^2}{v c_{\rm m}}\left[ \mathrm{Ci}\left(\frac{L_\| \omega_\mathrm{IR}}{v}\right) -  \mathrm{Ci}\left(\frac{L_\| \omega_\mathrm{UV}}{v}\right)+ \log \left(\frac{\omega_\mathrm{UV}}{\omega_\mathrm{IR}}\right)\right]} + O(T^0)\nonumber
\end{align}
where $\mathrm{Ci}(z) := \int_z^\infty \frac{-\mathrm{cos}(t)}{t} \mathrm{d}t$ is the cosine integral. 
In the $T \to \infty$ limit, the relevant timescale is instead given by $\tilde{t} \sim \frac{1}{T}$, hence the figure is plotted as a function of $\tilde{t}T$. 

In general, the probability distribution has the form:
\begin{align}\label{eq:delta}
&P(\Delta k_\|,v) =P_\mathrm{bounded}(\Delta k_\|,v) +   \\
&e^{ -\tilde{t} \frac{v^2 - c_m^2}{vc_m}(\omega_\mathrm{IR}-\omega_\mathrm{UV})} \left[ \frac{e^{\omega_\mathrm{UV}/T}-1}{e^{\omega_\mathrm{IR}/T}-1}\right]^{-2\tilde{t}T \frac{v^2 - c_{\rm m}^2}{vc_{\rm m}}} \delta(\Delta k_\|)\nonumber
\end{align}
where $P_\mathrm{bounded}(\Delta k_\|,v)$ is a bounded function of $\Delta k_\|$. 
As the IR-cutoff is removed $\omega_\mathrm{IR} \to 0$, the component proportional to $\delta(\Delta k_\|)$ vanishes for $T > 0$ and the resulting distributions $P(\Delta k_\|,v)$ are bounded functions for all $t > 0$. In all cases, after a long time has passed, the central limit theorem pulls the bulk of the distribution into a gaussian shape, where $P(\Delta {\bf k};{\bf v})$ asymptotically has mean $tM_i$ and covariance $tM_{ij}$. The tails of the distribution will remain non-Gaussian, consistent with the late-time limit~\eqref{eq:late-time}.

The discontinuous features appearing in \Cref{fig:fig} are associated with the hard cutoffs used in the regulator function, as well as the `boson-by-boson' nature of the emission. 
\Cref{eq:Sfinal} can be understood as the particle waiting a time according to an exponential distribution (see Eq.~\eqref{eq:delta}), before emitting/absorbing a gauge boson  ad infinitum. 
This is particularly clear in the $T = 0$ plots: the first emission corresponds to a uniform emission in the range $\Delta k_\| \in [\omega_\mathrm{IR}/v,\omega_\mathrm{UV}/v]$.  {Mathematically, this is a consequence of the fact that we can write
\begin{equation}
    P(\Delta {\bf k}; {\bf v} ) = \int \frac{\mathrm{d}^3 {\bf L}}{(2\pi)^3} e^{-i \Delta {\bf k} \cdot {\bf L} } \prod_{\omega,\phi} e^{- t S_{\omega,\phi}({\bf L};{\bf v}) } \, ,
\end{equation}
which is exactly the structure one gets for the distribution of a random variable $\Delta {\bf k}$ that is the sum of independent random variables in terms of their characteristic functions. As such, it is explicit that gauge bosons with different momenta are emitted independently, with rates determined by $S_{\omega,\phi}$ (the integrand in Eq.~\eqref{eq:Sfinal}).}

\textbf{Conclusions.}
The classical calculation of the Frank-Tamm formula provides the average energy lost due to a heavy charged particle travelling through a medium: quantum mechanics however asserts that this energy loss must occur due to the aggregate emission/absorption of individual photons. 
Our calculation presents the full quantum statistics associated to this emission, allowing us to explicitly describe the \textit{distribution} of the heavy particle's momentum, as well as the  \textit{distribution} of the gauge bosons emitted in the process per unit of frequency as an immediate corollary of energy conservation, directly from a straightforward QFT calculation.

Furthermore, our calculation was performed in the presence of a thermal background - allowing us to explicitly examine the thermal and quantum fluctuations of the energy loss. As evidenced by Figure~\ref{fig:fig}, our calculation demonstrates that while thermal effects can be understood straightforwardly in terms of stimulated emission and absorption at the microscopic level, their effect in the distribution of energy lost by the particle is quite complicated. This could potentially be of interest for heavy quark energy loss in QGP --- not as a realistic model, but rather as an explicit instance of how energy loss could conceivably take place, that serves to develop intuition in terms of momentum change probability distributions.

{In principle, $1/M$ corrections can be calculated by including the $O(1/M)$ pieces of the HQET Lagrangian. 
Whether or not the leading expression derived in this paper is appropriate depends on the comparison between the mass of the heavy charged particle and the intrinsic mass scales associated to the medium, which in our exposition are encoded within the cutoffs $\omega_\mathrm{UV}, \omega_\mathrm{IR}$. In actuality, the physical effect of the regulator function $\kappa(\omega)$ is induced by the frequency dependence of the permittivity and permeability of the medium ($\epsilon(\omega),\mu(\omega)$), ultimately leading to a frequency dependence of $c_\mathrm{med}(\omega)$. For example, for atmospheric Cherenkov radiation sourced by gamma ray showers~\cite{CTAConsortium:2010umy,CTAbook,Oya:2026amx}, the fact that the energy of the emitted photons is $\sim$eV and the energy of the electrons in the shower is certainly above $\sim$MeV means that the aforementioned corrections are rather small (similar for Cherenkov radiation in particle detectors at accelerators). As such, the dominant corrections to this result in realistic situations come from a more careful account of the properties of the medium (i.e., of $D^>$ in our formulas) and the showering process, which can be systematically examined starting from our formulas.
}

\acknowledgements

\textit{Acknowledgements.} This research was supported in part by grant NSF PHY-2309135 to the Kavli Institute for Theoretical Physics (KITP). BSH's contributions were supported by grant 994312 from the Simons Foundation.
JL's contributions were supported by the U.S. Department of Energy, Office of Science, Office of Nuclear Physics, Early Career Award through Contract No.~DE-SCL0000017.

\bibliographystyle{apsrev4-2}
\bibliography{draft.bib}

\begin{thebibliography}{51}%
\makeatletter
\providecommand \@ifxundefined [1]{%
 \@ifx{#1\undefined}
}%
\providecommand \@ifnum [1]{%
 \ifnum #1\expandafter \@firstoftwo
 \else \expandafter \@secondoftwo
 \fi
}%
\providecommand \@ifx [1]{%
 \ifx #1\expandafter \@firstoftwo
 \else \expandafter \@secondoftwo
 \fi
}%
\providecommand \natexlab [1]{#1}%
\providecommand \enquote  [1]{``#1''}%
\providecommand \bibnamefont  [1]{#1}%
\providecommand \bibfnamefont [1]{#1}%
\providecommand \citenamefont [1]{#1}%
\providecommand \href@noop [0]{\@secondoftwo}%
\providecommand \href [0]{\begingroup \@sanitize@url \@href}%
\providecommand \@href[1]{\@@startlink{#1}\@@href}%
\providecommand \@@href[1]{\endgroup#1\@@endlink}%
\providecommand \@sanitize@url [0]{\catcode `\\12\catcode `\$12\catcode `\&12\catcode `\#12\catcode `\^12\catcode `\_12\catcode `\%12\relax}%
\providecommand \@@startlink[1]{}%
\providecommand \@@endlink[0]{}%
\providecommand \url  [0]{\begingroup\@sanitize@url \@url }%
\providecommand \@url [1]{\endgroup\@href {#1}{\urlprefix }}%
\providecommand \urlprefix  [0]{URL }%
\providecommand \Eprint [0]{\href }%
\providecommand \doibase [0]{https://doi.org/}%
\providecommand \selectlanguage [0]{\@gobble}%
\providecommand \bibinfo  [0]{\@secondoftwo}%
\providecommand \bibfield  [0]{\@secondoftwo}%
\providecommand \translation [1]{[#1]}%
\providecommand \BibitemOpen [0]{}%
\providecommand \bibitemStop [0]{}%
\providecommand \bibitemNoStop [0]{.\EOS\space}%
\providecommand \EOS [0]{\spacefactor3000\relax}%
\providecommand \BibitemShut  [1]{\csname bibitem#1\endcsname}%
\let\auto@bib@innerbib\@empty
\bibitem [{\citenamefont {Frank}\ and\ \citenamefont {Tamm}(1937)}]{Frank:1937fk}%
  \BibitemOpen
  \bibfield  {author} {\bibinfo {author} {\bibfnamefont {I.~M.}\ \bibnamefont {Frank}}\ and\ \bibinfo {author} {\bibfnamefont {I.~E.}\ \bibnamefont {Tamm}},\ }\href {https://doi.org/10.3367/UFNr.0093.196710o.0388} {\bibfield  {journal} {\bibinfo  {journal} {Compt. Rend. Acad. Sci. URSS}\ }\textbf {\bibinfo {volume} {14}},\ \bibinfo {pages} {109} (\bibinfo {year} {1937})}\BibitemShut {NoStop}%
\bibitem [{\citenamefont {Bohr}(1948)}]{bohr1948penetration}%
  \BibitemOpen
  \bibfield  {author} {\bibinfo {author} {\bibfnamefont {N.}~\bibnamefont {Bohr}},\ }\href@noop {} {\emph {\bibinfo {title} {The Penetration of Atomic Particles Through Matter}}},\ \bibinfo {series} {Matematisk-fysiske meddelelser}, Vol.~\bibinfo {volume} {18}\ (\bibinfo  {publisher} {I kommission hos E. Munksgaard},\ \bibinfo {address} {Copenhagen},\ \bibinfo {year} {1948})\BibitemShut {NoStop}%
\bibitem [{\citenamefont {Jackson}(1998)}]{Jackson:1998nia}%
  \BibitemOpen
  \bibfield  {author} {\bibinfo {author} {\bibfnamefont {J.~D.}\ \bibnamefont {Jackson}},\ }\href@noop {} {\emph {\bibinfo {title} {{Classical Electrodynamics}}}}\ (\bibinfo  {publisher} {Wiley},\ \bibinfo {year} {1998})\BibitemShut {NoStop}%
\bibitem [{\citenamefont {Ginzburg}(1996)}]{Ginzburg_1996}%
  \BibitemOpen
  \bibfield  {author} {\bibinfo {author} {\bibfnamefont {V.~L.}\ \bibnamefont {Ginzburg}},\ }\href {https://doi.org/10.1070/PU1996v039n10ABEH000171} {\bibfield  {journal} {\bibinfo  {journal} {Physics-Uspekhi}\ }\textbf {\bibinfo {volume} {39}},\ \bibinfo {pages} {973} (\bibinfo {year} {1996})}\BibitemShut {NoStop}%
\bibitem [{met(1961)}]{methods-detection}%
  \BibitemOpen
  in\ \href {https://doi.org/https://doi.org/10.1016/S0076-695X(08)60442-1} {\emph {\bibinfo {booktitle} {Nuclear Physics}}},\ \bibinfo {series} {Methods in Experimental Physics}, Vol.~\bibinfo {volume} {5},\ \bibinfo {editor} {edited by\ \bibinfo {editor} {\bibfnamefont {L.~C.}\ \bibnamefont {Yuan}}\ and\ \bibinfo {editor} {\bibfnamefont {C.-S.}\ \bibnamefont {Wu}}}\ (\bibinfo  {publisher} {Academic Press},\ \bibinfo {year} {1961})\ pp.\ \bibinfo {pages} {1--288}\BibitemShut {NoStop}%
\bibitem [{\citenamefont {Agostinelli}\ \emph {et~al.}(2003)\citenamefont {Agostinelli} \emph {et~al.}}]{GEANT4:2002zbu}%
  \BibitemOpen
  \bibfield  {author} {\bibinfo {author} {\bibfnamefont {S.}~\bibnamefont {Agostinelli}} \emph {et~al.} (\bibinfo {collaboration} {GEANT4}),\ }\href {https://doi.org/10.1016/S0168-9002(03)01368-8} {\bibfield  {journal} {\bibinfo  {journal} {Nucl. Instrum. Meth. A}\ }\textbf {\bibinfo {volume} {506}},\ \bibinfo {pages} {250} (\bibinfo {year} {2003})}\BibitemShut {NoStop}%
\bibitem [{\citenamefont {Wada}(1949)}]{Wada:1949zz}%
  \BibitemOpen
  \bibfield  {author} {\bibinfo {author} {\bibfnamefont {W.~W.}\ \bibnamefont {Wada}},\ }\href {https://doi.org/10.1103/PhysRev.75.981} {\bibfield  {journal} {\bibinfo  {journal} {Phys. Rev.}\ }\textbf {\bibinfo {volume} {75}},\ \bibinfo {pages} {981} (\bibinfo {year} {1949})}\BibitemShut {NoStop}%
\bibitem [{\citenamefont {Ivanenko}\ and\ \citenamefont {Gurgenidze}(1949)}]{Ivanenko1949}%
  \BibitemOpen
  \bibfield  {author} {\bibinfo {author} {\bibfnamefont {D.~D.}\ \bibnamefont {Ivanenko}}\ and\ \bibinfo {author} {\bibfnamefont {V.~A.}\ \bibnamefont {Gurgenidze}},\ }\href@noop {} {\bibfield  {journal} {\bibinfo  {journal} {Dokl. Akad. Nauk SSSR}\ }\textbf {\bibinfo {volume} {67}},\ \bibinfo {pages} {997} (\bibinfo {year} {1949})}\BibitemShut {NoStop}%
\bibitem [{\citenamefont {Blokhintsev}\ and\ \citenamefont {Indenbom}(1950)}]{Blokhintsev1950}%
  \BibitemOpen
  \bibfield  {author} {\bibinfo {author} {\bibfnamefont {D.~I.}\ \bibnamefont {Blokhintsev}}\ and\ \bibinfo {author} {\bibfnamefont {V.~L.}\ \bibnamefont {Indenbom}},\ }\href@noop {} {\bibfield  {journal} {\bibinfo  {journal} {Zh. Eksp. Teor. Fiz.}\ }\textbf {\bibinfo {volume} {20}},\ \bibinfo {pages} {1123} (\bibinfo {year} {1950})}\BibitemShut {NoStop}%
\bibitem [{\citenamefont {Yekutieli}(1959)}]{Yekutieli1959}%
  \BibitemOpen
  \bibfield  {author} {\bibinfo {author} {\bibfnamefont {G.}~\bibnamefont {Yekutieli}},\ }\href@noop {} {\bibfield  {journal} {\bibinfo  {journal} {Nuovo Cim.}\ }\textbf {\bibinfo {volume} {13}},\ \bibinfo {pages} {446} (\bibinfo {year} {1959})}\BibitemShut {NoStop}%
\bibitem [{\citenamefont {Czyz}\ \emph {et~al.}(1959)\citenamefont {Czyz}, \citenamefont {Ericson},\ and\ \citenamefont {Glashow}}]{Czyz1959}%
  \BibitemOpen
  \bibfield  {author} {\bibinfo {author} {\bibfnamefont {W.}~\bibnamefont {Czyz}}, \bibinfo {author} {\bibfnamefont {T.}~\bibnamefont {Ericson}},\ and\ \bibinfo {author} {\bibfnamefont {S.~L.}\ \bibnamefont {Glashow}},\ }\href@noop {} {\bibfield  {journal} {\bibinfo  {journal} {Nucl. Phys.}\ }\textbf {\bibinfo {volume} {13}},\ \bibinfo {pages} {516} (\bibinfo {year} {1959})}\BibitemShut {NoStop}%
\bibitem [{\citenamefont {Czyz}\ and\ \citenamefont {Glashow}(1960)}]{Czyz1960}%
  \BibitemOpen
  \bibfield  {author} {\bibinfo {author} {\bibfnamefont {W.}~\bibnamefont {Czyz}}\ and\ \bibinfo {author} {\bibfnamefont {S.~L.}\ \bibnamefont {Glashow}},\ }\href@noop {} {\bibfield  {journal} {\bibinfo  {journal} {Nucl. Phys.}\ }\textbf {\bibinfo {volume} {20}},\ \bibinfo {pages} {309} (\bibinfo {year} {1960})}\BibitemShut {NoStop}%
\bibitem [{\citenamefont {Smrz}(1962)}]{Smrz1962}%
  \BibitemOpen
  \bibfield  {author} {\bibinfo {author} {\bibfnamefont {P.}~\bibnamefont {Smrz}},\ }\href@noop {} {\bibfield  {journal} {\bibinfo  {journal} {Nucl. Phys.}\ }\textbf {\bibinfo {volume} {35}},\ \bibinfo {pages} {165} (\bibinfo {year} {1962})}\BibitemShut {NoStop}%
\bibitem [{\citenamefont {Ion}(1970)}]{Ion1970}%
  \BibitemOpen
  \bibfield  {author} {\bibinfo {author} {\bibfnamefont {D.~B.}\ \bibnamefont {Ion}},\ }\href@noop {} {\bibfield  {journal} {\bibinfo  {journal} {St. Cerc. Fiz.}\ }\textbf {\bibinfo {volume} {22}},\ \bibinfo {pages} {125} (\bibinfo {year} {1970})}\BibitemShut {NoStop}%
\bibitem [{\citenamefont {Ion}\ and\ \citenamefont {Nichitiu}(1971)}]{Ion1971}%
  \BibitemOpen
  \bibfield  {author} {\bibinfo {author} {\bibfnamefont {D.~B.}\ \bibnamefont {Ion}}\ and\ \bibinfo {author} {\bibfnamefont {F.~G.}\ \bibnamefont {Nichitiu}},\ }\href@noop {} {\bibfield  {journal} {\bibinfo  {journal} {Nucl. Phys. B}\ }\textbf {\bibinfo {volume} {29}},\ \bibinfo {pages} {547} (\bibinfo {year} {1971})}\BibitemShut {NoStop}%
\bibitem [{\citenamefont {Ion}(1971)}]{Ion1971thesis}%
  \BibitemOpen
  \bibfield  {author} {\bibinfo {author} {\bibfnamefont {D.~B.}\ \bibnamefont {Ion}},\ }\emph {\bibinfo {title} {Mesonic Cerenkov-like Effect as Possible Mechanism of Meson Production in Hadronic Interactions}},\ \href@noop {} {\bibinfo {type} {Dsc thesis}},\ \bibinfo  {school} {Bucharest University} (\bibinfo {year} {1971})\BibitemShut {NoStop}%
\bibitem [{\citenamefont {Zaretskii}\ and\ \citenamefont {Lomonosov}(1977)}]{Zaretskii1977}%
  \BibitemOpen
  \bibfield  {author} {\bibinfo {author} {\bibfnamefont {D.~F.}\ \bibnamefont {Zaretskii}}\ and\ \bibinfo {author} {\bibfnamefont {V.~V.}\ \bibnamefont {Lomonosov}},\ }\href@noop {} {\bibfield  {journal} {\bibinfo  {journal} {Sov. J. Nucl. Phys.}\ }\textbf {\bibinfo {volume} {26}},\ \bibinfo {pages} {639} (\bibinfo {year} {1977})}\BibitemShut {NoStop}%
\bibitem [{\citenamefont {Dremin}(1979)}]{Dremin1979}%
  \BibitemOpen
  \bibfield  {author} {\bibinfo {author} {\bibfnamefont {I.~M.}\ \bibnamefont {Dremin}},\ }\href@noop {} {\bibfield  {journal} {\bibinfo  {journal} {JETP Lett.}\ }\textbf {\bibinfo {volume} {30}},\ \bibinfo {pages} {140} (\bibinfo {year} {1979})},\ \bibinfo {note} {pisma v ZhETF {\bf 30} (1979) 152}\BibitemShut {NoStop}%
\bibitem [{\citenamefont {Dremin}(1981)}]{Dremin1981}%
  \BibitemOpen
  \bibfield  {author} {\bibinfo {author} {\bibfnamefont {I.~M.}\ \bibnamefont {Dremin}},\ }\href@noop {} {\bibfield  {journal} {\bibinfo  {journal} {Sov. J. Nucl. Phys.}\ }\textbf {\bibinfo {volume} {33}},\ \bibinfo {pages} {726} (\bibinfo {year} {1981})},\ \bibinfo {note} {yad. Fiz. {\bf 33} (1981) 1357}\BibitemShut {NoStop}%
\bibitem [{\citenamefont {Dremin}(2006)}]{Dremin:2005an}%
  \BibitemOpen
  \bibfield  {author} {\bibinfo {author} {\bibfnamefont {I.~M.}\ \bibnamefont {Dremin}},\ }\href {https://doi.org/10.1016/j.nuclphysa.2005.12.015} {\bibfield  {journal} {\bibinfo  {journal} {Nucl. Phys. A}\ }\textbf {\bibinfo {volume} {767}},\ \bibinfo {pages} {233} (\bibinfo {year} {2006})},\ \Eprint {https://arxiv.org/abs/hep-ph/0507167} {arXiv:hep-ph/0507167} \BibitemShut {NoStop}%
\bibitem [{\citenamefont {Price}\ \emph {et~al.}(2026)\citenamefont {Price}, \citenamefont {S.~Formanek},\ and\ \citenamefont {Rafelski}}]{Price:2026mof}%
  \BibitemOpen
  \bibfield  {author} {\bibinfo {author} {\bibfnamefont {W.}~\bibnamefont {Price}}, \bibinfo {author} {\bibfnamefont {M.}~\bibnamefont {S.~Formanek}},\ and\ \bibinfo {author} {\bibfnamefont {J.}~\bibnamefont {Rafelski}},\ }\href@noop {} {\  (\bibinfo {year} {2026})},\ \Eprint {https://arxiv.org/abs/2603.01636} {arXiv:2603.01636 [hep-ph]} \BibitemShut {NoStop}%
\bibitem [{\citenamefont {Kaufhold}\ and\ \citenamefont {Klinkhamer}(2006)}]{Kaufhold:2005vj}%
  \BibitemOpen
  \bibfield  {author} {\bibinfo {author} {\bibfnamefont {C.}~\bibnamefont {Kaufhold}}\ and\ \bibinfo {author} {\bibfnamefont {F.~R.}\ \bibnamefont {Klinkhamer}},\ }\href {https://doi.org/10.1016/j.nuclphysb.2005.11.001} {\bibfield  {journal} {\bibinfo  {journal} {Nucl. Phys. B}\ }\textbf {\bibinfo {volume} {734}},\ \bibinfo {pages} {1} (\bibinfo {year} {2006})},\ \Eprint {https://arxiv.org/abs/hep-th/0508074} {arXiv:hep-th/0508074} \BibitemShut {NoStop}%
\bibitem [{\citenamefont {Altschul}(2007)}]{Altschul:2006zz}%
  \BibitemOpen
  \bibfield  {author} {\bibinfo {author} {\bibfnamefont {B.}~\bibnamefont {Altschul}},\ }\href {https://doi.org/10.1103/PhysRevLett.98.041603} {\bibfield  {journal} {\bibinfo  {journal} {Phys. Rev. Lett.}\ }\textbf {\bibinfo {volume} {98}},\ \bibinfo {pages} {041603} (\bibinfo {year} {2007})},\ \Eprint {https://arxiv.org/abs/hep-th/0609030} {arXiv:hep-th/0609030} \BibitemShut {NoStop}%
\bibitem [{\citenamefont {Cheng}\ \emph {et~al.}(2006)\citenamefont {Cheng}, \citenamefont {Luty}, \citenamefont {Mukohyama},\ and\ \citenamefont {Thaler}}]{Cheng:2006us}%
  \BibitemOpen
  \bibfield  {author} {\bibinfo {author} {\bibfnamefont {H.-C.}\ \bibnamefont {Cheng}}, \bibinfo {author} {\bibfnamefont {M.~A.}\ \bibnamefont {Luty}}, \bibinfo {author} {\bibfnamefont {S.}~\bibnamefont {Mukohyama}},\ and\ \bibinfo {author} {\bibfnamefont {J.}~\bibnamefont {Thaler}},\ }\href {https://doi.org/10.1088/1126-6708/2006/05/076} {\bibfield  {journal} {\bibinfo  {journal} {JHEP}\ }\textbf {\bibinfo {volume} {05}},\ \bibinfo {pages} {076}},\ \Eprint {https://arxiv.org/abs/hep-th/0603010} {arXiv:hep-th/0603010} \BibitemShut {NoStop}%
\bibitem [{\citenamefont {Carroll}\ \emph {et~al.}(2009)\citenamefont {Carroll}, \citenamefont {Tam},\ and\ \citenamefont {Wehus}}]{Carroll:2009mr}%
  \BibitemOpen
  \bibfield  {author} {\bibinfo {author} {\bibfnamefont {S.~M.}\ \bibnamefont {Carroll}}, \bibinfo {author} {\bibfnamefont {H.}~\bibnamefont {Tam}},\ and\ \bibinfo {author} {\bibfnamefont {I.~K.}\ \bibnamefont {Wehus}},\ }\href {https://doi.org/10.1103/PhysRevD.80.025020} {\bibfield  {journal} {\bibinfo  {journal} {Phys. Rev. D}\ }\textbf {\bibinfo {volume} {80}},\ \bibinfo {pages} {025020} (\bibinfo {year} {2009})},\ \Eprint {https://arxiv.org/abs/0904.4680} {arXiv:0904.4680 [hep-th]} \BibitemShut {NoStop}%
\bibitem [{\citenamefont {De~Laurentis}\ \emph {et~al.}(2012)\citenamefont {De~Laurentis}, \citenamefont {Capozziello},\ and\ \citenamefont {Basini}}]{DeLaurentis:2012fh}%
  \BibitemOpen
  \bibfield  {author} {\bibinfo {author} {\bibfnamefont {M.}~\bibnamefont {De~Laurentis}}, \bibinfo {author} {\bibfnamefont {S.}~\bibnamefont {Capozziello}},\ and\ \bibinfo {author} {\bibfnamefont {G.}~\bibnamefont {Basini}},\ }\href {https://doi.org/10.1142/S0217732312501362} {\bibfield  {journal} {\bibinfo  {journal} {Mod. Phys. Lett. A}\ }\textbf {\bibinfo {volume} {27}},\ \bibinfo {pages} {1250136} (\bibinfo {year} {2012})},\ \Eprint {https://arxiv.org/abs/1206.6681} {arXiv:1206.6681 [gr-qc]} \BibitemShut {NoStop}%
\bibitem [{\citenamefont {Kimura}\ and\ \citenamefont {Yamamoto}(2012)}]{Kimura:2011qn}%
  \BibitemOpen
  \bibfield  {author} {\bibinfo {author} {\bibfnamefont {R.}~\bibnamefont {Kimura}}\ and\ \bibinfo {author} {\bibfnamefont {K.}~\bibnamefont {Yamamoto}},\ }\href {https://doi.org/10.1088/1475-7516/2012/07/050} {\bibfield  {journal} {\bibinfo  {journal} {JCAP}\ }\textbf {\bibinfo {volume} {07}},\ \bibinfo {pages} {050}},\ \Eprint {https://arxiv.org/abs/1112.4284} {arXiv:1112.4284 [astro-ph.CO]} \BibitemShut {NoStop}%
\bibitem [{\citenamefont {Yagi}\ \emph {et~al.}(2014{\natexlab{a}})\citenamefont {Yagi}, \citenamefont {Blas}, \citenamefont {Barausse},\ and\ \citenamefont {Yunes}}]{Yagi:2013ava}%
  \BibitemOpen
  \bibfield  {author} {\bibinfo {author} {\bibfnamefont {K.}~\bibnamefont {Yagi}}, \bibinfo {author} {\bibfnamefont {D.}~\bibnamefont {Blas}}, \bibinfo {author} {\bibfnamefont {E.}~\bibnamefont {Barausse}},\ and\ \bibinfo {author} {\bibfnamefont {N.}~\bibnamefont {Yunes}},\ }\href {https://doi.org/10.1103/PhysRevD.89.084067} {\bibfield  {journal} {\bibinfo  {journal} {Phys. Rev. D}\ }\textbf {\bibinfo {volume} {89}},\ \bibinfo {pages} {084067} (\bibinfo {year} {2014}{\natexlab{a}})},\ \bibinfo {note} {[Erratum: Phys.Rev.D 90, 069902 (2014), Erratum: Phys.Rev.D 90, 069901 (2014)]},\ \Eprint {https://arxiv.org/abs/1311.7144} {arXiv:1311.7144 [gr-qc]} \BibitemShut {NoStop}%
\bibitem [{\citenamefont {Yagi}\ \emph {et~al.}(2014{\natexlab{b}})\citenamefont {Yagi}, \citenamefont {Blas}, \citenamefont {Yunes},\ and\ \citenamefont {Barausse}}]{Yagi:2013qpa}%
  \BibitemOpen
  \bibfield  {author} {\bibinfo {author} {\bibfnamefont {K.}~\bibnamefont {Yagi}}, \bibinfo {author} {\bibfnamefont {D.}~\bibnamefont {Blas}}, \bibinfo {author} {\bibfnamefont {N.}~\bibnamefont {Yunes}},\ and\ \bibinfo {author} {\bibfnamefont {E.}~\bibnamefont {Barausse}},\ }\href {https://doi.org/10.1103/PhysRevLett.112.161101} {\bibfield  {journal} {\bibinfo  {journal} {Phys. Rev. Lett.}\ }\textbf {\bibinfo {volume} {112}},\ \bibinfo {pages} {161101} (\bibinfo {year} {2014}{\natexlab{b}})},\ \Eprint {https://arxiv.org/abs/1307.6219} {arXiv:1307.6219 [gr-qc]} \BibitemShut {NoStop}%
\bibitem [{\citenamefont {Yunes}\ \emph {et~al.}(2016)\citenamefont {Yunes}, \citenamefont {Yagi},\ and\ \citenamefont {Pretorius}}]{Yunes:2016jcc}%
  \BibitemOpen
  \bibfield  {author} {\bibinfo {author} {\bibfnamefont {N.}~\bibnamefont {Yunes}}, \bibinfo {author} {\bibfnamefont {K.}~\bibnamefont {Yagi}},\ and\ \bibinfo {author} {\bibfnamefont {F.}~\bibnamefont {Pretorius}},\ }\href {https://doi.org/10.1103/PhysRevD.94.084002} {\bibfield  {journal} {\bibinfo  {journal} {Phys. Rev. D}\ }\textbf {\bibinfo {volume} {94}},\ \bibinfo {pages} {084002} (\bibinfo {year} {2016})},\ \Eprint {https://arxiv.org/abs/1603.08955} {arXiv:1603.08955 [gr-qc]} \BibitemShut {NoStop}%
\bibitem [{\citenamefont {Rajagopal}\ \emph {et~al.}(2025{\natexlab{a}})\citenamefont {Rajagopal}, \citenamefont {Scheihing-Hitschfeld},\ and\ \citenamefont {Wiedemann}}]{Rajagopal:2025ukd}%
  \BibitemOpen
  \bibfield  {author} {\bibinfo {author} {\bibfnamefont {K.}~\bibnamefont {Rajagopal}}, \bibinfo {author} {\bibfnamefont {B.}~\bibnamefont {Scheihing-Hitschfeld}},\ and\ \bibinfo {author} {\bibfnamefont {U.~A.}\ \bibnamefont {Wiedemann}},\ }\href {https://doi.org/10.1007/JHEP07(2025)013} {\bibfield  {journal} {\bibinfo  {journal} {JHEP}\ }\textbf {\bibinfo {volume} {07}},\ \bibinfo {pages} {013}},\ \Eprint {https://arxiv.org/abs/2501.06289} {arXiv:2501.06289 [hep-ph]} \BibitemShut {NoStop}%
\bibitem [{\citenamefont {Rajagopal}\ \emph {et~al.}(2025{\natexlab{b}})\citenamefont {Rajagopal}, \citenamefont {Scheihing-Hitschfeld},\ and\ \citenamefont {Wiedemann}}]{Rajagopal:2025rxr}%
  \BibitemOpen
  \bibfield  {author} {\bibinfo {author} {\bibfnamefont {K.}~\bibnamefont {Rajagopal}}, \bibinfo {author} {\bibfnamefont {B.}~\bibnamefont {Scheihing-Hitschfeld}},\ and\ \bibinfo {author} {\bibfnamefont {U.~A.}\ \bibnamefont {Wiedemann}},\ }\href {https://doi.org/10.1103/m2d2-9sdb} {\bibfield  {journal} {\bibinfo  {journal} {Phys. Rev. Lett.}\ }\textbf {\bibinfo {volume} {135}},\ \bibinfo {pages} {242301} (\bibinfo {year} {2025}{\natexlab{b}})},\ \Eprint {https://arxiv.org/abs/2504.21139} {arXiv:2504.21139 [hep-ph]} \BibitemShut {NoStop}%
\bibitem [{\citenamefont {Sirunyan}\ \emph {et~al.}(2018)\citenamefont {Sirunyan} \emph {et~al.}}]{CMS:2017qjw}%
  \BibitemOpen
  \bibfield  {author} {\bibinfo {author} {\bibfnamefont {A.~M.}\ \bibnamefont {Sirunyan}} \emph {et~al.} (\bibinfo {collaboration} {CMS}),\ }\href {https://doi.org/10.1016/j.physletb.2018.05.074} {\bibfield  {journal} {\bibinfo  {journal} {Phys. Lett. B}\ }\textbf {\bibinfo {volume} {782}},\ \bibinfo {pages} {474} (\bibinfo {year} {2018})},\ \Eprint {https://arxiv.org/abs/1708.04962} {arXiv:1708.04962 [nucl-ex]} \BibitemShut {NoStop}%
\bibitem [{\citenamefont {Acharya}\ \emph {et~al.}(2018)\citenamefont {Acharya} \emph {et~al.}}]{ALICE:2018lyv}%
  \BibitemOpen
  \bibfield  {author} {\bibinfo {author} {\bibfnamefont {S.}~\bibnamefont {Acharya}} \emph {et~al.} (\bibinfo {collaboration} {ALICE}),\ }\href {https://doi.org/10.1007/JHEP10(2018)174} {\bibfield  {journal} {\bibinfo  {journal} {JHEP}\ }\textbf {\bibinfo {volume} {10}},\ \bibinfo {pages} {174}},\ \Eprint {https://arxiv.org/abs/1804.09083} {arXiv:1804.09083 [nucl-ex]} \BibitemShut {NoStop}%
\bibitem [{\citenamefont {Acharya}\ \emph {et~al.}(2021)\citenamefont {Acharya} \emph {et~al.}}]{ALICE:2021mgk}%
  \BibitemOpen
  \bibfield  {author} {\bibinfo {author} {\bibfnamefont {S.}~\bibnamefont {Acharya}} \emph {et~al.} (\bibinfo {collaboration} {ALICE}),\ }\href {https://doi.org/10.1007/JHEP05(2021)220} {\bibfield  {journal} {\bibinfo  {journal} {JHEP}\ }\textbf {\bibinfo {volume} {05}},\ \bibinfo {pages} {220}},\ \Eprint {https://arxiv.org/abs/2102.13601} {arXiv:2102.13601 [nucl-ex]} \BibitemShut {NoStop}%
\bibitem [{\citenamefont {Acharya}\ \emph {et~al.}(2022)\citenamefont {Acharya} \emph {et~al.}}]{ALICE:2021rxa}%
  \BibitemOpen
  \bibfield  {author} {\bibinfo {author} {\bibfnamefont {S.}~\bibnamefont {Acharya}} \emph {et~al.} (\bibinfo {collaboration} {ALICE}),\ }\href {https://doi.org/10.1007/JHEP01(2022)174} {\bibfield  {journal} {\bibinfo  {journal} {JHEP}\ }\textbf {\bibinfo {volume} {01}},\ \bibinfo {pages} {174}},\ \Eprint {https://arxiv.org/abs/2110.09420} {arXiv:2110.09420 [nucl-ex]} \BibitemShut {NoStop}%
\bibitem [{\citenamefont {Abdulameer}\ \emph {et~al.}(2024)\citenamefont {Abdulameer} \emph {et~al.}}]{PHENIX:2022wim}%
  \BibitemOpen
  \bibfield  {author} {\bibinfo {author} {\bibfnamefont {N.~J.}\ \bibnamefont {Abdulameer}} \emph {et~al.} (\bibinfo {collaboration} {PHENIX}),\ }\href {https://doi.org/10.1103/PhysRevC.109.044907} {\bibfield  {journal} {\bibinfo  {journal} {Phys. Rev. C}\ }\textbf {\bibinfo {volume} {109}},\ \bibinfo {pages} {044907} (\bibinfo {year} {2024})},\ \Eprint {https://arxiv.org/abs/2203.17058} {arXiv:2203.17058 [nucl-ex]} \BibitemShut {NoStop}%
\bibitem [{\citenamefont {Hayrapetyan}\ \emph {et~al.}(2025)\citenamefont {Hayrapetyan} \emph {et~al.}}]{CMS:2024vip}%
  \BibitemOpen
  \bibfield  {author} {\bibinfo {author} {\bibfnamefont {A.}~\bibnamefont {Hayrapetyan}} \emph {et~al.} (\bibinfo {collaboration} {CMS}),\ }\href {https://doi.org/10.1007/JHEP02(2025)195} {\bibfield  {journal} {\bibinfo  {journal} {JHEP}\ }\textbf {\bibinfo {volume} {02}},\ \bibinfo {pages} {195}},\ \Eprint {https://arxiv.org/abs/2409.07258} {arXiv:2409.07258 [nucl-ex]} \BibitemShut {NoStop}%
\bibitem [{\citenamefont {Baier}\ \emph {et~al.}(2000)\citenamefont {Baier}, \citenamefont {Schiff},\ and\ \citenamefont {Zakharov}}]{Baier:2000mf}%
  \BibitemOpen
  \bibfield  {author} {\bibinfo {author} {\bibfnamefont {R.}~\bibnamefont {Baier}}, \bibinfo {author} {\bibfnamefont {D.}~\bibnamefont {Schiff}},\ and\ \bibinfo {author} {\bibfnamefont {B.~G.}\ \bibnamefont {Zakharov}},\ }\href {https://doi.org/10.1146/annurev.nucl.50.1.37} {\bibfield  {journal} {\bibinfo  {journal} {Ann. Rev. Nucl. Part. Sci.}\ }\textbf {\bibinfo {volume} {50}},\ \bibinfo {pages} {37} (\bibinfo {year} {2000})},\ \Eprint {https://arxiv.org/abs/hep-ph/0002198} {arXiv:hep-ph/0002198} \BibitemShut {NoStop}%
\bibitem [{\citenamefont {Casalderrey-Solana}\ and\ \citenamefont {Salgado}(2007)}]{Casalderrey-Solana:2007knd}%
  \BibitemOpen
  \bibfield  {author} {\bibinfo {author} {\bibfnamefont {J.}~\bibnamefont {Casalderrey-Solana}}\ and\ \bibinfo {author} {\bibfnamefont {C.~A.}\ \bibnamefont {Salgado}},\ }\href@noop {} {\bibfield  {journal} {\bibinfo  {journal} {Acta Phys. Polon. B}\ }\textbf {\bibinfo {volume} {38}},\ \bibinfo {pages} {3731} (\bibinfo {year} {2007})},\ \Eprint {https://arxiv.org/abs/0712.3443} {arXiv:0712.3443 [hep-ph]} \BibitemShut {NoStop}%
\bibitem [{\citenamefont {Beraudo}\ \emph {et~al.}(2018)\citenamefont {Beraudo} \emph {et~al.}}]{Rapp:2018qla}%
  \BibitemOpen
  \bibfield  {author} {\bibinfo {author} {\bibfnamefont {A.}~\bibnamefont {Beraudo}} \emph {et~al.},\ }\href {https://doi.org/10.1016/j.nuclphysa.2018.09.002} {\bibfield  {journal} {\bibinfo  {journal} {Nucl. Phys. A}\ }\textbf {\bibinfo {volume} {979}},\ \bibinfo {pages} {21} (\bibinfo {year} {2018})},\ \Eprint {https://arxiv.org/abs/1803.03824} {arXiv:1803.03824 [nucl-th]} \BibitemShut {NoStop}%
\bibitem [{\citenamefont {Wang}\ and\ \citenamefont {Wiedemann}(2025)}]{Wang:2025lct}%
  \BibitemOpen
  \bibfield  {author} {\bibinfo {author} {\bibfnamefont {X.-N.}\ \bibnamefont {Wang}}\ and\ \bibinfo {author} {\bibfnamefont {U.~A.}\ \bibnamefont {Wiedemann}}\ }(\bibinfo {year} {2025})\ \Eprint {https://arxiv.org/abs/2508.18794} {arXiv:2508.18794 [hep-ph]} \BibitemShut {NoStop}%
\bibitem [{\citenamefont {Koch}\ \emph {et~al.}(2006)\citenamefont {Koch}, \citenamefont {Majumder},\ and\ \citenamefont {Wang}}]{Koch:2005sx}%
  \BibitemOpen
  \bibfield  {author} {\bibinfo {author} {\bibfnamefont {V.}~\bibnamefont {Koch}}, \bibinfo {author} {\bibfnamefont {A.}~\bibnamefont {Majumder}},\ and\ \bibinfo {author} {\bibfnamefont {X.-N.}\ \bibnamefont {Wang}},\ }\href {https://doi.org/10.1103/PhysRevLett.96.172302} {\bibfield  {journal} {\bibinfo  {journal} {Phys. Rev. Lett.}\ }\textbf {\bibinfo {volume} {96}},\ \bibinfo {pages} {172302} (\bibinfo {year} {2006})},\ \Eprint {https://arxiv.org/abs/nucl-th/0507063} {arXiv:nucl-th/0507063} \BibitemShut {NoStop}%
\bibitem [{\citenamefont {Casalderrey-Solana}\ \emph {et~al.}(2010{\natexlab{a}})\citenamefont {Casalderrey-Solana}, \citenamefont {Fernandez},\ and\ \citenamefont {Mateos}}]{Casalderrey-Solana:2009zbk}%
  \BibitemOpen
  \bibfield  {author} {\bibinfo {author} {\bibfnamefont {J.}~\bibnamefont {Casalderrey-Solana}}, \bibinfo {author} {\bibfnamefont {D.}~\bibnamefont {Fernandez}},\ and\ \bibinfo {author} {\bibfnamefont {D.}~\bibnamefont {Mateos}},\ }\href {https://doi.org/10.1103/PhysRevLett.104.172301} {\bibfield  {journal} {\bibinfo  {journal} {Phys. Rev. Lett.}\ }\textbf {\bibinfo {volume} {104}},\ \bibinfo {pages} {172301} (\bibinfo {year} {2010}{\natexlab{a}})},\ \Eprint {https://arxiv.org/abs/0912.3717} {arXiv:0912.3717 [hep-ph]} \BibitemShut {NoStop}%
\bibitem [{\citenamefont {Casalderrey-Solana}\ \emph {et~al.}(2010{\natexlab{b}})\citenamefont {Casalderrey-Solana}, \citenamefont {Fernandez},\ and\ \citenamefont {Mateos}}]{Casalderrey-Solana:2010cfu}%
  \BibitemOpen
  \bibfield  {author} {\bibinfo {author} {\bibfnamefont {J.}~\bibnamefont {Casalderrey-Solana}}, \bibinfo {author} {\bibfnamefont {D.}~\bibnamefont {Fernandez}},\ and\ \bibinfo {author} {\bibfnamefont {D.}~\bibnamefont {Mateos}},\ }\href {https://doi.org/10.1007/JHEP11(2010)091} {\bibfield  {journal} {\bibinfo  {journal} {JHEP}\ }\textbf {\bibinfo {volume} {11}},\ \bibinfo {pages} {091}},\ \Eprint {https://arxiv.org/abs/1009.5937} {arXiv:1009.5937 [hep-th]} \BibitemShut {NoStop}%
\bibitem [{\citenamefont {Georgi}(1990)}]{Georgi:1990um}%
  \BibitemOpen
  \bibfield  {author} {\bibinfo {author} {\bibfnamefont {H.}~\bibnamefont {Georgi}},\ }\href {https://doi.org/10.1016/0370-2693(90)91128-X} {\bibfield  {journal} {\bibinfo  {journal} {Phys. Lett. B}\ }\textbf {\bibinfo {volume} {240}},\ \bibinfo {pages} {447} (\bibinfo {year} {1990})}\BibitemShut {NoStop}%
\bibitem [{\citenamefont {Kramers}(1940)}]{Kramers1940}%
  \BibitemOpen
  \bibfield  {author} {\bibinfo {author} {\bibfnamefont {H.}~\bibnamefont {Kramers}},\ }\href {https://doi.org/https://doi.org/10.1016/S0031-8914(40)90098-2} {\bibfield  {journal} {\bibinfo  {journal} {Physica}\ }\textbf {\bibinfo {volume} {7}},\ \bibinfo {pages} {284} (\bibinfo {year} {1940})}\BibitemShut {NoStop}%
\bibitem [{\citenamefont {Moyal}(1949)}]{Moyal1949}%
  \BibitemOpen
  \bibfield  {author} {\bibinfo {author} {\bibfnamefont {J.~E.}\ \bibnamefont {Moyal}},\ }\href {https://doi.org/10.1111/j.2517-6161.1949.tb00030.x} {\bibfield  {journal} {\bibinfo  {journal} {Journal of the Royal Statistical Society: Series B (Methodological)}\ }\textbf {\bibinfo {volume} {11}},\ \bibinfo {pages} {150} (\bibinfo {year} {1949})}\BibitemShut {NoStop}%
\bibitem [{\citenamefont {Actis}\ \emph {et~al.}(2011)\citenamefont {Actis} \emph {et~al.}}]{CTAConsortium:2010umy}%
  \BibitemOpen
  \bibfield  {author} {\bibinfo {author} {\bibfnamefont {M.}~\bibnamefont {Actis}} \emph {et~al.} (\bibinfo {collaboration} {CTA Consortium}),\ }\href {https://doi.org/10.1007/s10686-011-9247-0} {\bibfield  {journal} {\bibinfo  {journal} {Exper. Astron.}\ }\textbf {\bibinfo {volume} {32}},\ \bibinfo {pages} {193} (\bibinfo {year} {2011})},\ \Eprint {https://arxiv.org/abs/1008.3703} {arXiv:1008.3703 [astro-ph.IM]} \BibitemShut {NoStop}%
\bibitem [{\citenamefont {CTAConsortium}(2019)}]{CTAbook}%
  \BibitemOpen
  \bibfield  {author} {\bibinfo {author} {\bibnamefont {CTAConsortium}},\ }\href {https://doi.org/10.1142/10986} {\emph {\bibinfo {title} {{Science with the Cherenkov Telescope Array}}}}\ (\bibinfo  {publisher} {WORLD SCIENTIFIC},\ \bibinfo {year} {2019})\ \Eprint {https://arxiv.org/abs/https://www.worldscientific.com/doi/pdf/10.1142/10986} {https://www.worldscientific.com/doi/pdf/10.1142/10986} \BibitemShut {NoStop}%
\bibitem [{\citenamefont {Oya}(2026)}]{Oya:2026amx}%
  \BibitemOpen
  \bibfield  {author} {\bibinfo {author} {\bibfnamefont {I.}~\bibnamefont {Oya}} (\bibinfo {collaboration} {CTAO}),\ }\href {https://doi.org/10.1016/j.nima.2026.171414} {\bibfield  {journal} {\bibinfo  {journal} {Nucl. Instrum. Meth. A}\ }\textbf {\bibinfo {volume} {1087}},\ \bibinfo {pages} {171414} (\bibinfo {year} {2026})}\BibitemShut {NoStop}%
\end{thebibliography}%

\appendix

\onecolumngrid

\section{Derivation in SI units}

Here we discuss the formulas introduced in the main text for the specific case of Quantum Electrodynamics, in the SI unit system, for a readier comparison with textbook electromagnetism.

The Maxwell action in a linear medium with permittivity  $\epsilon$ and magnetic permeability $\mu$ can be written as
\begin{equation}
    S_{\rm Maxwell} = \frac12 \int \mathrm{d}t \ \mathrm{d}^3 x \left[ \epsilon {\bf E}^2 - \frac{1}{\mu} {\bf B}^2 \right] \, , 
\end{equation}
where the speed of light in the medium is
\begin{equation}
    c_{\rm m}^2 = \frac{1}{\mu \epsilon} \, ,
\end{equation}
and the electric and magnetic fields are given by
\begin{align}
    {\bf E} = \frac{\partial {\bf A} }{\partial t} - \nabla A_0 \, , & &
    {\bf B} = \nabla \times {\bf A} \, .
\end{align}

Like any gauge theory, this action has a gauge freedom $A_\mu \to A_\mu + \partial_\mu \Lambda$. In order to have a path integral formulation of the quantum theory, a gauge must be chosen. Here we add a gauge-fixing term to the Maxwell action
\begin{equation}
    \Delta S_{\rm gauge-fixing} = -\frac{\epsilon}{2c_{\rm m}^2} \int \mathrm{d}t \ \mathrm{d}^3x \left( \partial_t A_0 - c_{\rm m}^2 \partial_i A_i \right)^2 \, ,
\end{equation}
which corresponds to Feynman gauge. The resulting action is
\begin{equation}
    S = \frac{\epsilon}{2} \int \mathrm{d}t \ \mathrm{d}^3x \left[ - \left( \frac{1}{c_{\rm m}^2} (\partial_t A_0)^2 - (\partial_i A_0)(\partial_i A_0) \right) + \left( (\partial_t A_i )(\partial_t A_i ) - c_{\rm m}^2 (\partial_i A_j) (\partial_i A_j) \right) \right] \, ,
\end{equation}
which yields a momentum space Wightman function
\begin{align}
    D^>_{\mu \nu}(\omega,{\bf k}) &= - \hbar \epsilon^{-1} c_{\rm m}^2 g_{\mu \nu}\frac{ {\rm sgn}(\omega)  }{1- e^{-\hbar \omega/T}} (2\pi) \delta(\omega^2 - c_{\rm m}^2 {\bf k}^2 ) \, ,
\end{align}
with $g_{\mu \nu} = {\rm diag}(1,-c_{\rm m}^{-2},-c_{\rm m}^{-2},-c_{\rm m}^{-2})$.

With $\hbar$ restored, the Wilson loop derived from a point charge $q$ moving in a straight line reads
\begin{align} 
    \langle W_{\bf v} \rangle ({\bf L}) &= \exp \left( t q^2 \hbar^{-2} n^\mu n^\nu \int_{-\infty}^\infty dt' \left[ D_{\mu \nu}^{>}(t',{\bf v} t' + {\bf L} ) -  D_{\mu \nu}^{>}(t',{\bf v} t' ) \right] \right) \nonumber \\
    &= \exp \left( t \frac{q^2  (v^2 - c_{\rm m}^2 ) }{8 \pi^2 \hbar \epsilon v c_{\rm m}^2 }  \int_{-\infty}^{+\infty} \mathrm{d}\omega \int_0^{2\pi} \mathrm{d}\phi \frac{(e^{i L_\perp k_\perp \cos \phi } e^{i \omega L_\|/v}-1)\ \mathrm{sgn}(\omega)}{1 - e^{-\hbar \omega/T}} \right) \nonumber \\
    &= \exp \left( t \frac{\mu q^2}{4\pi \hbar}  v \left(1 - \frac{c_{\rm m}^2}{v^2} \right)   \int_{-\infty}^{+\infty} \mathrm{d}\omega \int_0^{2\pi} \frac{\mathrm{d}\phi}{2\pi} \frac{(e^{i L_\perp k_\perp \cos \phi } e^{i \omega L_\|/v}-1)\ \mathrm{sgn}(\omega)}{1 - e^{-\hbar \omega/T}} \right) \, . 
\end{align}

The radiated energy loss rate is given by (noting that the momentum change is $\hbar {\bf k}$)
\begin{align}
    \frac{\mathrm{d}E}{\mathrm{d}x} &=  \frac{{\bf v}_i}{v} M_i = \frac{1}{vt}  {\bf v} \cdot (-i \hbar \nabla) \ln \langle W_{\bf v} \rangle ({\bf L}) \, ,
\end{align}
which yields, in the zero temperature limit
\begin{equation}
    \frac{\partial^2 E}{\partial x \partial \omega} = \frac{q^2}{4\pi} \mu \left(1 - \frac{c_{\rm m}^2}{v^2} \right) \, ,
\end{equation}
the Frank-Tamm formula in SI units.

The fact that the fluctuations around this result are quantum-mechanical in nature is clear by noting that
\begin{align}
    P(\Delta {\bf p} = \hbar \Delta {\bf k};{\bf v}) &= \frac{1}{(2\pi)^3} \int \mathrm{d}^3{\bf L} \, e^{-i \Delta {\bf p} \cdot {\bf L} /\hbar  } \langle W_{\bf v} \rangle ({\bf L}) \, , \label{eq:P-from-W-hbar}
\end{align}
which means that
\begin{equation}
    \int \mathrm{d}^3 (\Delta {\bf p}) \,  P(\Delta {\bf p};{\bf v}) (\Delta {\bf p}_{i_1}) \cdots (\Delta {\bf p}_{i_n}) = (-i \hbar \partial_{i_1} ) \cdots (-i \hbar \partial_{i_n} ) \left. \langle W_{\bf v} \rangle ({\bf L}) \right|_{{\bf L} = 0 } \, ,
\end{equation}
i.e., an additional factor of $\hbar$ is generated with every higher moment. The mean is $O(\hbar^{0})$, the variance is $O(\hbar)$, and so on. 

\end{document}